\documentstyle[amssymb,aps]{revtex}

\begin{document}

\twocolumn[\hsize\textwidth\columnwidth\hsize \csname @twocolumnfalse\endcsname             
\title{{\bf Nuclear isotope thermometry}}
\author{ S.R. Souza$^{a}$, W.P. Tan$^{b}$,  R. Donangelo$^{a}$,  C.K. Gelbke$^{b}$,  W.G. Lynch$^{b}$  and M.B. Tsang$^{b}$}
\address{$^{a}$Instituto de Fisica, Universidade Federal do Rio de Janeiro\\
Cidade Universit\'aria, CP 68528, 21945-970 Rio de Janeiro, Brazil\\
$^{b}$Department of Physics and Astronomy and National Superconducting Cyclotron
Laboratory, Michigan State University,\\
East Lansing, Michigan 48824 }
\maketitle

\begin{abstract}
We discuss different aspects which could influence temperatures
deduced from experimental isotopic yields in the multifragmentation
process.
It is shown that fluctuations due to
the finite size of the system and distortions due to
the decay of hot primary fragments conspire to blur the temperature
determination in multifragmentation reactions.
These facts suggest that caloric curves obtained
through isotope thermometers, which were taken as
evidence for a first-order phase transition in nuclear matter,
should be investigated very carefully.
\end{abstract}

\pacs{24.60.-k, 25.70.Pq, 25.70.-z}

]

\narrowtext

\section{Introduction\newline
}

\label{sec:introduction} Due to the short range attraction between nucleons,
nuclear matter is a Fermi liquid \cite{Baym} at low temperature and is
expected to undergo a phase transition to a nucleonic gas within a mixed
phase region bounded by a critical temperature of order 15 MeV \cite
{Jaqaman,muller}. Experimental investigations of this phase transition have
focused on a variety of experimental observables ranging from the mass,
charge or multiplicity distributions for the emitted fragments \cite
{williams,schuttauf} to observables sensitive to the temperature of the
system \cite{benenson,Pochodzalla}.

Temperature measurements, in particular, have been performed to search for
evidence of the enhanced heat capacity predicted by statistical model
calculations reflecting the latent heat for transforming the Fermi liquid to
the nucleonic vapor \cite{benenson,Pochodzalla,huang}. For example, the
Statistical Multifragmentation Model (SMM) \cite{Bondorf} predicts a plateau
of roughly constant temperature of $T\approx 5MeV$ for excitation energies
of $E^{*}/A\approx 3-7MeV.$ At these excitation energies the model predicts
a mixed phase consisting of fragments (liquid) and nucleons and light
particles (gas) corresponding to a mixed phase equilibrium. This is followed
at higher excitation energies by a linear rise in the temperature with
excitation energy as expected for a gas of small nuclei having negligible
internal heat capacity \cite{Bondorf}. Similar effects are predicted by the
Microcanonical Metropolis Monte Carlo (MMMC) model \cite{gross}.

This trend was qualitatively reproduced in some experiments \cite
{Pochodzalla}, but not in others \cite{ma,hauger,serfling,Xi98}. An
essential part of these measurements is the determination of the temperature
of the fragmenting system. Temperatures were extracted from the isotopic
abundances of helium and lithium fragments, using the isotope thermometry
method proposed by Albergo {\it et al.} \cite{Albergo}. The idea of the
method is to determine the double ratios of the yields of four suitably
chosen isotopes, ($A_{1},Z_{1}$), ($A_{1}+1,Z_{1}$), ($A_{2},Z_{2}$), ($%
A_{2}+1,Z_{2}$),

\begin{equation}
\frac{Y(A_{1},Z_{1})/Y(A_{1}+1,Z_{1})}{Y(A_{2},Z_{2})/Y(A_{2}+1.Z_{2})}%
=C\exp (\Delta B/T_{iso})  \label{eq:t_alb}
\end{equation}

\noindent where the $Y$ are the yields of the different isotopes, $C$ is a
constant related to spin values and kinematic factors, $\Delta
B=B(A_{1},Z_{1})-B(A_{1}+1,Z_{1})-B(A_{2},Z_{2})+B(A_{2}+1,Z_{2})$ is
obtained from the binding energies of the isotopes appearing in Eq.~(\ref
{eq:t_alb}), and $T_{iso}$ stands for the temperature deduced from this
isotopic thermometer. In the case of the He-Li thermometer employed in ref.~%
\cite{Pochodzalla}, $A_{1}=6,Z_{1}=3,A_{2}=3,Z_{2}=2$. For the C-Li
thermometer, more recently considered by Xi {\it et al.} \cite{Xi98}, $%
A_{1}=6,Z_{1}=3,A_{2}=11,Z_{2}=6$. For the Carbon thermometer studied in
this work, $A_{1}=12,Z_{1}=6,A_{2}=11,Z_{2}=6.$

However, there are a few aspects which should be carefully analyzed when one
wants to compare information on the breakup configuration of an excited
system formed in a heavy-ion collision to multifragmentation models like the
SMM approach. Some of these points are addressed below. In sect.\ \ref
{sec:assumptions} we briefly discuss the assumptions underlying this method.
Variations in the temperature of the breakup stage, where the hot primary
fragments decouple from the system, are intrinsic to finite systems and are
explored within the SMM approach in sect. \ref{sec:prim_y}. An analytic
description of temperature variations is developed in the grand canonical
limit in sect.\ \ref{sec:tempfluct}; this description is consistent with the
results from the SMM. In addition, there are finite size effects, discussed
in sect. \ref{sec:chempot}, that make the concept of an overall chemical
potential somewhat inaccurate. The influence of secondary decay is discussed
in sect. \ref{sec:sec_dec}. Conclusions are drawn in sect.\ \ref{sec:conrem}.

\section{Underlying assumptions\newline
}

\label{sec:assumptions} The basic physical hypotheses of the isotope
thermometry method are:

\begin{enumerate}
\item  an equilibrated source is formed after the most violent stages of the
reaction and it decays simultaneously and statistically,

\item  for the experimental event selection employed in the analyses, all
the events correspond to fragments formed at the same temperature, and

\item  distortions on the isotopic temperature due to secondary decay of hot
primary fragments may be neglected.
\end{enumerate}

\noindent Although the Statistical Multifragmentation Model \cite{Bondorf},
used in the discussion below, is based on the first assumption, the last two
hypotheses are not supported by the model, as we shall discuss in detail.

The SMM uses the Monte Carlo method and averages observables with the
statistical weight over decay partitions. A multifragment decay partition is 
{\it defined} in the SMM approach \cite{Bondorf} as a specific set of
emitted fragments and light particles. For simplicity, each partition in the
SMM approach is weighted according to the entropy of the partition. This
entropy is approximated by analytical expressions rather than by an event by
event sampling of the phase space as in ref. \cite{gross} These
approximations rely upon that fact that the dominant contribution to this
entropy comes from internal phase space of fragments which plays the role of
a heat bath within the SMM approach just as an excited residue plays the
role of a heat bath within compound nuclear decay theory \cite{friedman}.

For a given decay partition and by making a Wigner Seitz approximation to
the Coulomb energy, energy conservation within the SMM approach leads to the
expression \cite{Bondorf},

\begin{equation}
E_{0}^{gs}+E_{0}^{*}=\frac{3}{5}\frac{Z_{0}^{2}e^{2}}{R_{0}}+\sum_{\left\{
A,Z\right\} }N_{AZ}E_{AZ}  \label{eq:econs}
\end{equation}

\noindent where $E_{0}^{*}$ is the total excitation energy and $E_{0}^{gs}$
is ground state energy of a nuclei having a mass and atomic number equal to
that of the total system, $A_{0}$ and $Z_{0}$, respectively. The first term
on the right hand side stands for the Coulomb energy of a homogeneous charge 
$Z_{0}e$ occupying the volume of the system of radius $R_{0}$ and $N_{AZ}$
indicates the number of fragments of mass number $A$ and atomic number $Z$
in the partition of the system.

In the equation above, $E_{AZ}$ is the kinetic plus internal energy for each
of these fragments. It is related to the temperature by assuming all
fragments are at a common temperature as follows,

\begin{equation}
E_{AZ}=\frac{3}{2}T+E_{AZ}^*(T)+E_{AZ}^C-B_{AZ}  \label{eq:eaz}
\end{equation}

\noindent where the internal excitation energy of the fragments, $%
E_{AZ}^{*}(T)$, may be approximated by an extension of the semi-empirical
mass formula to finite temperatures \cite{Bondorf}, and the extra Coulomb
energy of the fragment in the fragmentation volume, $E_{AZ}^{C}$, may be
calculated within the Wigner-Seitz approximation. $B_{AZ}$ stands for the
ground state binding energy for the fragment. Eqs. \ref{eq:econs} and \ref
{eq:eaz} result from an average of the microcanonical expression for energy
conservation over the phase space corresponding to the specific decay
partition.

By applying the energy conservation relationship in Eqs. (\ref{eq:econs}-\ref
{eq:eaz}) one obtains a temperature $T$ that describes the internal
excitation and translational energies of fragments within a given partition.
Even though the overall system is assumed to be in equilibrium at a fixed
excitation energy $E_{0}^{*},$ different decay partitions have different
Coulomb, binding, and translational energies and, consequently, different
excitation energies of the emitted fragments. Consistency with Eqs. (\ref
{eq:econs}-\ref{eq:eaz}) therefore requires that the temperature $T$ of the
fragments varies from one decay partition to another, reflecting the
differences between the Coulomb, binding and translational energies of the
various partitions.

Labeling the partition $\{N_{AZ}\}$ with the index, $f,$ the statistical
weight associated with the partition,

\begin{equation}
W_{f}=\exp \left[ \sum_{\left\{ A,Z\right\} }N_{AZ}S_{AZ}(T)\right] \,
\label{SAZ}
\end{equation}

\noindent may be found by expressing the entropy of the fragments, $S_{AZ}$,
using approximations derived from the liquid drop model at finite
temperature \cite{Bondorf}. Consequently the physical observables can be
expressed by a weighted average over decay partitions as,

\begin{equation}
\left\langle O_{AZ}\right\rangle =\frac{\sum_{f}W_{f}O_{AZ}}{\sum_{f}W_{f}}
\label{eq:ob}
\end{equation}
where $O_{AZ}$ can be any interesting observables such as the yield of a
fragment or the temperature (In the present work, the summation included $%
10^{8}$ partitions.)

This allows one to predict the various results from the SMM that are
addressed in the next section with regard to the temperature
variations.Because the SMM approach invokes a temperature to sample the
microcanonical phase space, we denote the predicted observables as {\it %
approximately} microcanonical. Despite this caveat, we note that this
procedure can, in principal, give accurate microcanonical predictions for
experimental observables provided the thermal expressions for the free
energies are accurate descriptions of the integration over the
microcanonical phase space.

Before passing on to the various results of our investigation, it is
important to clarify that we do not invoke the Grand Canonical
approximations to the SMM approach introduced in ref. \cite{Botvina87} to
allow Monte Carlo event simulations\cite{foot1}. Instead, we have adhered
closely to original SMM approach outlined in ref. \cite{Bondorf}, with the
exception that all calculations in this paper were performed at a constant
freezeout density equal to one third that of the saturation density of
nuclear matter.

\section{Primary Temperatures}

\label{sec:prim_y} The SMM procedure expressed in Eqs.(\ref{eq:econs}-\ref
{eq:ob}) leads to a distribution of the temperatures of the fragmenting
system for a given excitation energy in the same sense that the temperature
of the daughter nucleus in compound nuclear decay theory varies as a
function of the Coulomb barrier and separation energy of each decay channel.
The points in Fig.~(\ref{Tdist}) denote the temperature distributions for
the fragmentation of an excited $^{112}Sn$ nucleus at three different
excitation energies obtained with the SMM. These distributions are well
fitted by gaussian functions, shown by the curves in the figure, with
variances $\sigma _{T}^{2}$ that are fairly independent of the energy, $%
\sigma _{T}\approx 0.4$ MeV, in the range $3\,{\rm MeV}\,\le E_{0}^{*}/A\le
10\,{\rm MeV}$. At each excitation energy, we average over all of the
partitions and define this average value as the '' approximate
microcanonical'' temperature T$_{MIC}.$

Since each of the isotopes employed in the thermometer has a specific mass,
charge and binding energy, the application of conservation laws sets a
constraint on the values available to the remainder of the system. Because
of this finite size effect, the temperature distribution obtained when a
specific isotope is present is slightly different from the one obtained when
all partitions are considered. In particular, a small difference ($\leq 0.1$
MeV) is observed between the average temperatures for the various isotopes;
this is illustrated in Fig.~(\ref{Tdist_i}) for carbon isotopes from the
fragmentation of a $^{112}Sn$ nucleus at $E_{0}^{*}/A=6$ MeV. Even though
the average temperatures are different reflecting the different binding
energies of the three isotopes, all these distributions are gaussians with
nearly the same variances. We can extract another temperature $T_{IMF\text{ }%
}$ by averaging over partitions which contain an Intermediate Mass Fragment
(IMF) with $3\leq Z\leq 10$. It's interesting to note that $T_{MIC}$ can
exceed $T_{IMF}$ at low energies by as much as 0.2 MeV, in part because it
takes more energy to emit an IMF than to emit an equivalent mass in the form
of alpha particles, leaving less energy for thermal excitation.

The basic idea contained in Eq.\ (\ref{eq:t_alb}) was derived under the
assumption that the primary yields are well represented by the grand
canonical approximation at a single breakup temperature; the double ratio
was invoked to cancel out the contribution to the yields coming from the
neutron and proton chemical potentials. In the SMM, however, the temperature
varies from partition to partition and the chemical potentials, which appear
within the grand canonical formalism as Lagrange multipliers that conserve
charge and mass, are not explicitly invoked. Thus, we can not presume the
validity of the Albergo's formula ( Eq.\ \ref{eq:t_alb}) in the SMM and must
test its validity instead.

We begin with a test of the validity of Eq.\ (\ref{eq:t_alb}) when one
employs the primary yields. For a given decay partition $\{N_{AZ}\},$ we
take into account the internal free energy $F_{AZ}^{\text{int}}(T)$
\noindent which is parameterized as:

\begin{eqnarray}
F_{AZ}^{\text{int}} &=&-B(A,Z)+F_{AZ}^{*}(T)+F_{AZ}^{C}\;,  \label{eq:efree}
\\
F_{AZ}^{*}(T) &=&F_{AZ}^{B*}(T)+F_{AZ}^{S*}(T)-T\ln (g_{AZ}^{gs})
\label{eq:estar}
\end{eqnarray}

\noindent where $g_{AZ}^{gs}$ is the ground state spin degeneracy, and $%
F_{AZ}^{B*}$, $F_{AZ}^{S*}$, and $F_{AZ}^{C}$ correspond to the excitation
energy dependent bulk, surface, and Coulomb contributions to the internal
free energy\cite{foot} after the binding energy part has been removed. The
reader is referred to ref.\ \cite{Bondorf} for explicit expressions for the
terms entering in the equation above. Then the primary yield for the ground
state can be related to the total yield by

\begin{equation}
N_{AZ}^{gs}=N_{AZ}\cdot g_{AZ}^{gs}\exp \left[ F_{AZ}^{*}(T)/T\right]
\end{equation}
for this partition. Following the procedure described in the previous
section, we will use this expression and Eq. (\ref{eq:ob}) to obtain the
average g.s. yield distribution $\left\langle N_{AZ}^{gs}\right\rangle $.
This, in turn, can be used in Eq. (\ref{eq:t_alb}) to extract isotopic
temperatures as follows,

\begin{equation}
\frac{\left\langle N_{A1,Z1}^{gs}\right\rangle /\left\langle
N_{A1+1,Z1}^{gs}\right\rangle }{\left\langle N_{A2,Z2}^{gs}\right\rangle
/\left\langle N_{A2+1,Z2}^{gs}\right\rangle }=C\exp \left( \frac{\Delta B}{%
T_{iso}^{smm}}\right) .  \label{eq:tprim}
\end{equation}
In previous SMM calculations, experimental binding energies and spin
degeneracy factors $g_{AZ}^{gs}$ were used for light nuclei with $A<5$.
Liquid drop binding energies and spin degeneracy factors of unity were used
for $A\ge 5$. \ In this work, we will retain these conventions on spin
degeneracy factors so as to be consistent with prior calculations, but we
will use empirical binding energies for all nuclei.

In Fig. (\ref{Thermo}), the isotopic temperatures $T_{iso}^{smm}$ for the
carbon thermometer ($Z_{1}=Z_{2}=6,A_{1}=11,A_{2}=12$) are plotted as the
stars for the multifragmentation of a $^{112}Sn$ source at excitation
energies $E_{0}^{*}/A=3-10$ MeV$.$ For comparisons, the corresponding $%
T_{MIC}$ and $T_{IMF}$ for the same system are also shown in Fig.\ (\ref
{Thermo}) as the dashed and solid lines, respectively. While supporting the
concept of isotopic thermometry, the good agreement between $T_{IMF}$ and $%
T_{iso}^{smm}$ is somewhat surprising, given the strong dependence of the
Boltzmann factor on temperature for large $\Delta B$ and the width of the
temperature distribution shown in Fig.~(\ref{Tdist}). As shown in the
following section, it occurs in part due to a large cancellation involving
the Boltzmann factor and the temperature dependences of the effective
chemical potentials. Fig. (\ref{Thermo}) also reveals that fairly precise
information about $T_{IMF}$ and somewhat less precise information about $%
T_{MIC}$ is provided by the primary yields. This suggests that given a
precise relationship between primary to the final yields, it would be
possible to determine the breakup temperature from the measured yields.

\section{Effects of Temperature Variations}

\label{sec:tempfluct}The surprising consistency between $T_{IMF}$ and $%
T_{iso}^{smm}$ in Fig. (\ref{Thermo}) suggests that the corrections to the
grand canonical prediction for the isotope temperatures are small, and one
may utilize this approach to understand why the temperature variations have
so little influence on the results. Taking this tact, we assume that the
isotopic distributions are well approximated for each partition by the grand
canonical limit, use this limit to gain insight into the finite size effects
and at the same time, investigate the accuracy of this approximation. We
take this approach to consider first the influence of the temperature
variations and later the consequences of the finite size on the effective
chemical potentials.

Considering the influence of the temperature variations in this
approximation, we average the grand canonical approximation over the
temperature distribution in Fig.~(\ref{Tdist}). If the approximation works,
the expressions that result from this average should be appropriate for the
consideration of the effects of temperature distributions arising from other
effects and within other equilibrium models of multifragmentation as well.
Taking this approach, the yield of a particular isotope $i$ in the framework
of Albergo's method \cite{Albergo}, when averaged over all possible
partitions, becomes:

\begin{eqnarray}
&\langle Y_{i}\rangle =&V\int_{0}^{\infty }dTf(T){\frac{A_{i}^{3/2}\zeta
_{i}(T)}{\lambda _{T}^{3}}}  \nonumber \\
&&\exp \left[ (Z_{i}\,\mu _{PF}\left( T\right) +N_{i}\,\mu _{NF}\left(
T\right) +B_{i})/T\right] \;  \label{eq:Y_albf}
\end{eqnarray}

\noindent where $f(T)$ is the temperature distribution, $V$ represents the
free volume of the system, $\lambda _{T}=\sqrt{2\pi \hbar ^{2}/mT}$, $m$ is
the nucleon mass and $\mu _{PF}$ ($\mu _{NF}$) stands for the chemical
potential associated with free protons (neutrons) at temperature $T$. The
internal partition function of the fragment $i$ is given by:

\begin{equation}
\zeta _{i}(T)=\sum_{j}g_{i}^{j}\exp \left[ -{\frac{\Delta E_{j}}{T}}\right]
\;  \label{eq:int_pf}
\end{equation}

\noindent where $\Delta E_{j}$ is the excitation energy of the state $j$
with respect to the ground state and $g_{i}^{j}$ stands for the spin
degeneracy factor of this excited state.

Assuming that $f(T)$ is a gaussian centered at $\langle T\rangle $ and with
width $\sigma _{T}\ll \langle T\rangle $ (see Fig.~\ref{Tdist} ), one may
expand $1/T$, $T^{3/2}$, and the chemical potentials. By considering only
fragments observed in the ground state, {\it i.e.} $\zeta _{i}(T)=g_{i}^{0}$%
, we obtain that

\begin{eqnarray}
\langle Y_{i}^{gs}\rangle &=&{\frac{g_{i}^{0}\,V\,A_{i}^{3/2}\,\langle
T\rangle ^{3/2}}{\lambda _{*}^{3}}}  \nonumber \\
&&{\cdot \exp }\left[ \frac{B_{i}}{\left\langle T\right\rangle }+\frac{\mu
_{PF}\left( \left\langle T\right\rangle \right) Z_{i}+\mu _{NF}\left(
\left\langle T\right\rangle \right) N_{i}}{\left\langle T\right\rangle }%
\right] \;  \nonumber \\
&&\cdot \frac{1}{\sqrt{2p}}\cdot \exp \left[ \frac{q^{2}}{4p}\right] .
\label{eq:aveY}
\end{eqnarray}

\noindent where $\lambda _{*}\equiv \sqrt{2\pi \hbar ^{2}/m}$. In the above
expression, the corrections to the grand canonical relationship are provided
by the correction factor $\frac{1}{\sqrt{2p}}\cdot \exp \left[ \frac{q^{2}}{%
4p}\right] $ which depends on assumed width of the temperature distribution,
the binding energy of the i-th fragment, the neutron and proton chemical
potentials and their derivatives through the parameters $p$ and $q$. These
two parameters are defined by

\begin{eqnarray}
p &=&\frac{1}{2}+\left[ \frac{\sigma _{T}}{\left\langle T\right\rangle }%
\right] ^{2}\cdot \left[ Z_{i}\alpha _{PF}+N_{i}\alpha _{NF}+\frac{3}{4}-%
\frac{B_{i}}{\left\langle T\right\rangle }\right]  \label{eq:corrP} \\
q &=&\frac{\sigma _{T}}{\left\langle T\right\rangle }\left( Z_{i}\beta
_{PF}+N_{i}\beta _{NF}+\frac{3}{2}-\frac{B_{i}}{\left\langle T\right\rangle }%
\right) ,  \nonumber
\end{eqnarray}

\noindent where

\begin{eqnarray}
\alpha _{PF} &=&\mu _{PF}^{^{\prime }}\left( \left\langle T\right\rangle
\right) -\frac{\mu _{PF}\left( \left\langle T\right\rangle \right) }{%
\left\langle T\right\rangle }-\frac{1}{2}\mu _{PF}^{^{\prime \prime }}\left(
\left\langle T\right\rangle \right) \left\langle T\right\rangle
\label{eq:corrQ} \\
\beta _{PF} &=&\mu _{PF}^{^{\prime }}\left( \left\langle T\right\rangle
\right) -\frac{\mu _{PF}\left( \left\langle T\right\rangle \right) }{%
\left\langle T\right\rangle }  \nonumber \\
\alpha _{NF} &=&\mu _{NF}^{^{\prime }}\left( \left\langle T\right\rangle
\right) -\frac{\mu _{NF}\left( \left\langle T\right\rangle \right) }{%
\left\langle T\right\rangle }-\frac{1}{2}\mu _{NF}^{^{\prime \prime }}\left(
\left\langle T\right\rangle \right) \left\langle T\right\rangle  \nonumber \\
\beta _{NF} &=&\mu _{NF}^{^{\prime }}\left( \left\langle T\right\rangle
\right) -\frac{\mu _{NF}\left( \left\langle T\right\rangle \right) }{%
\left\langle T\right\rangle }.  \nonumber
\end{eqnarray}

The isotopic temperature can be extracted from the above corrected yields.
Replacing $Y(A,Z)$ in Eq.\ (\ref{eq:t_alb}) by the right hand side of Eq.\ (%
\ref{eq:aveY}), one cancels out the spin and mass dependent term C and then
obtains:

\begin{equation}
\exp \left[ \Delta B/T_{iso}^{cal}\right] ={\frac{%
G(A_{1},Z_{1})/G(A_{1}+1,Z_{1})}{G(A_{2},Z_{2})/G(A_{2}+1,Z_{2})}}\;,
\label{eq:cor_alb}
\end{equation}

\noindent where

\begin{eqnarray}
G(A,Z) &=&{\exp }\left[ \frac{B_{i}}{\left\langle T\right\rangle }+\frac{\mu
_{PF}\left( \left\langle T\right\rangle \right) Z+\mu _{NF}\left(
\left\langle T\right\rangle \right) N}{\left\langle T\right\rangle }\right]
\;  \nonumber \\
&&\cdot \frac{1}{\sqrt{2p}}\cdot \exp \left[ \frac{q^{2}}{4p}\right] \;.
\label{eq:g}
\end{eqnarray}
\noindent In the above double ratio the terms involving the chemical
potentials evaluated at the average temperature cancel; however, terms in
the correction factor involving the derivatives of the chemical potentials
remain.

Quantitative estimates of the correction factor require one to obtain
estimates for the effective chemical potentials and their derivatives with
respect to temperature. The proton and neutron chemical potentials at
temperature $T$ may be calculated from the free proton and neutron
multiplicities via the expression:

\begin{eqnarray}
\mu _{PF}(T) &=&T\log \left[ {\frac{\lambda _{T}^{3}Y_{PF}(T)}{g_{PF}V}}%
\right] \;,  \label{eq:chp} \\
\mu _{NF}(T) &=&T\log \left[ {\frac{\lambda _{T}^{3}Y_{NF}(T)}{g_{NF}V}}%
\right]  \nonumber
\end{eqnarray}

\noindent where $g_{PF}(g_{NF})$ represents the spin degeneracy factor of
the proton(neutron). For the calculations presented in this work, it has
proven advantageous and reasonably accurate to approximate the yields $%
Y_{PF}(T)$ and $Y_{NF}(T)$ over a modest range in temperature by power law
expressions in the temperature. In this approximation,

\begin{eqnarray}
Y_{PF}(T) &=&{c}_{PF}T^{\gamma _{PF}}\;,  \label{temp_expa} \\
Y_{NF}(T) &=&{c}_{NF}T^{\gamma _{NF}}  \nonumber
\end{eqnarray}
\noindent For the decay of $^{112}Sn$ nuclei at temperatures ranging over $%
4\leq T\leq 7MeV$ , $Y_{PF}$ and $Y_{NF}$ are well described by $\gamma
_{PF}=4.5$ and $\gamma _{NF}=1.0$ $(c_{PF}=1.33\times 10^{-4}$ and $%
c_{NF}=0.267)$ according to the SMM; comparisons of this parameterization to
yields calculated with the SMM model are shown Fig. (\ref{PN_fit})$.$ These
values depend on the density, which has been chosen to be one third that of
the saturation density of nuclear matter. Larger values of the free nucleon
yields are obtained at lower density.

Using this approximation, the explicit forms of the correction factors in
Eqs. (\ref{eq:aveY})-(\ref{eq:corrQ}) become 2$\alpha _{PF}=\beta
_{PF}=\left( \gamma _{PF}-\frac{3}{2}\right) =3$ and 2$\alpha _{NF}=\beta
_{NF}=\left( \gamma _{NF}-\frac{3}{2}\right) =-\frac{1}{2}$. We note that
the correction factor to the temperature $T_{iso}^{cal}$ in Eq.\ (\ref
{eq:cor_alb}) depends on the power law exponents $\gamma _{PF}(\gamma _{NF})$
and not on the absolute values of the proton(neutron) yields.

Even though Eq. (\ref{eq:Y_albf}) has an exponent that appears to be
strongly temperature dependent, there is a strong cancelation between the
contributions from the chemical potentials and binding energy factors in the
expressions for $p$ and $q.$ As a result, the correction factor is of order
unity. Values in the range of $\frac{1}{\sqrt{2p}}\cdot \exp \left[ \frac{%
q^{2}}{4p}\right] \approx 1-2$ are obtained, for example, in the decay of $%
^{112}Sn$ nuclei at temperatures in the range of $4\leq T\leq 7MeV$ .

The isotopic temperatures $T_{iso}^{cal}$ calculated from Eq.\ (\ref
{eq:cor_alb}) for carbon thermometer are shown in Fig. (\ref{Thermo}) in
comparisons with temperatures $T_{MIC}$, $T_{IMF}$ and $T_{iso}^{smm}$
derived from the SMM in the previous session. The very good agreement
between $T_{iso}^{cal}$ , $T_{iso}^{smm}$ and $T_{IMF}$ indicates that the
corrections to the isotopic temperatures associated with these temperature
variations are small, although the yields can change by as much as a factor
of two. This comparative insensitivity arises because the isotopic
thermometers depend logarithmically on the yields.

This insensitivity depends on the nature and magnitude of the temperature
variation. The corrections to the isotopic temperatures will be somewhat
larger in other contexts or other models where the temperature variations
are larger. The limited precision with which systems may be selected
experimentally may also have a similar influence because the excitation
energy and temperature varies experimentally from collision to collision due
to variations in the impact parameter or in the energy removed by
preequilibrium particle emission. The influence of this temperature
variation, which may exceed the variation in temperature caused by the
averaging over decay partitions, can also be estimated via techniques
outlined in the present section. To illustrate how one can estimate the
possible corrections due to an imprecision in the excitation energy
definition, the circles in Fig. (\ref{Thermo}) show calculations using Eq. (%
\ref{eq:cor_alb}) for carbon thermometer assuming a width of $\sigma
_{T}\approx 0.8$ MeV for the temperature distribution, which is twice as
large as that predicted in Figs. 1 and 2. This width is not based upon a
dynamical calculation; it is only to illustrate that larger isotopic
temperatures can result if the excitation energy is poorly defined.

\section{Chemical Potentials}

\label{sec:chempot}

The grand canonical limit has a great advantage of providing a simple
analytic expression for the isotopic yields from which other useful
expressions can be derived. However, the concept of uniform chemical
potentials is not a prediction of microcanonical or canonical models and
must be investigated to determine its applicability to finite systems. We do
this by trying to comparing the grand canonical expression for the isotopic
yields to the predictions of approximately microcanonical SMM calculations.
We start by assuming that these isotopic distributions can be calculated
within the grand canonical approximation and then test this assumption as
follows.Using a pair of adjacent isotopes, we invert the grand canonical
expression for the isotopic yields of two adjacent isotopes to obtain an
equation for the {\it effective }neutron chemical potential:

\begin{eqnarray}
\mu _{n}^{eff}\left( A,Z\right) &=&T\log [\frac{g_{AZ}^{gs}}{g_{A+1Z}^{gs}}%
\left( \frac{A}{A+1}\right) ^{3/2}  \nonumber \\
&&\exp \left( \left( B_{AZ}-B_{A+1Z}\right) /T\right) \frac{Y_{A+1Z}^{gs}}{%
Y_{AZ}^{gs}}]  \label{eq:n_effchem}
\end{eqnarray}
where g$_{AZ}^{gs}$, $B_{AZ}$ and $Y_{AZ}^{gs}$ are the ground state spin
degeneracy, the binding energy and the ground state primary yield for a
fragment with (A,Z), respectively. If the $Y_{AZ}^{gs}$ taken to be the
ground state yields predicted by the SMM, $\mu _{n}^{eff}\left( A,Z\right) $
becomes an effective ''SMM'' chemical potential. By performing SMM
calculations, we find the temperature- and isotopic- dependences of the
effective neutron chemical potentials given in Fig. (\ref{fig:n_chem}) for
Carbon and Lithium isotopes from the decay of a $^{112}$Sn nucleus at
excitation energies of $E_{0}^{*}/A=3,6,9\;MeV$.

These effective chemical potentials are essentially the same for the Carbon
and Lithium isotope chains. This insensitivity to element number is
consistent with the concept of a chemical potential and offers support for
the use of the grand canonical expression to describe isotopic
distributions. There is a dependence on the neutron number of the isotope,
however, that lies outside of the grand canonical approximation. This
variation in the neutron chemical potential basically comes as a result of
mass, charge and energy conservation for a finite-size system. We can
understand the influence of these conservation laws most easily at low
excitation energies, where the two largest fragments in the final state are
the IMF (Carbon or Lithium in this case) and a heavy residue which contains
most of the remaining charge and mass. We estimate the influence of
conservation laws at low excitation energy qualitatively by considering
binary decay configurations. Assuming that a parent nucleus ($A_{0},Z_{0}$)
decays into a light fragment (A,Z) and a heavy residue $(A_{0}-A,Z_{0}-Z)$ ,
we can approximate the yield of fragment (A,Z) in its ground state by

\begin{eqnarray}
Y_{AZ}^{gs} &\varpropto &\rho ^{gs}\left( A,Z\right) \rho ^{*}\left(
A_{0}-A,Z_{0}-Z\right) \overline{\rho }_{REL}  \label{eq:Y_bin} \\
&\thickapprox &g_{AZ}^{gs}\cdot \exp \left[ S^{*}\left(
A_{0}-A,Z_{0}-Z\right) \right]  \nonumber \\
&&\cdot \left[ \frac{A\cdot \left( A_{0}-A\right) }{A_{0}}\right] ^{3/2}%
\frac{1}{\lambda _{T}^{3}}  \nonumber
\end{eqnarray}
where $\rho ^{gs}=g_{AZ}^{gs},$ $\rho ^{*}$ and $S^{*}$ are the density of
states for the light nucleus in its ground state level, the density of
states and entropy of the heavy residue in its excited state, respectively,
The other factor, $\overline{\rho }_{REL}\thickapprox \left[ \frac{A\cdot
\left( A_{0}-A\right) }{A_{0}}\right] ^{3/2}\lambda _{T}^{-3},$ is the
thermal average of the state density of relative motion.

Replacing the yields in Eq.(\ref{eq:n_effchem}) with Eq.(\ref{eq:Y_bin}) and
assuming $A<<A_{0},$ one finds that the effective chemical potentiail
depends on the difference in residue entropies, $S^{*}\left( A_{0}-A-1,%
\overline{Z}\right) -S^{*}\left( A_{0}-A,Z_{0}-Z\right) $. Using an
expansion for small changes in the nuclear entropy from ref. \cite{friedman}%
, this difference can be expressed in terms of the difference of binding
energies, 
\begin{eqnarray}
&&S^{*}\left( A_{0}-A-1,\overline{Z}\right) -S^{*}\left(
A_{0}-A,Z_{0}-Z\right)  \nonumber \\
&=&-(B_{A_{0}-A,Z_{0}-Z}-B_{A_{0}-A-1,Z_{0}-Z})/T  \nonumber \\
&&-(B_{AZ}-B_{A+1Z})/T+f^{*}/T
\end{eqnarray}
plus a term depending on the free excitation energy per nucleon, $%
f^{*}=E^{*}/A_{0}-TS/A_{0}.$ This difference in binding energies is further
related to the neutron separation energy $s_{n}(A_{0}-A,Z_{0}-Z)$:

\begin{equation}
s_{n}(A_{0}-A,Z_{0}-Z)=B_{A_{0}-A,Z_{0}-Z}-B_{A_{0}-A-1,Z_{0}-Z}
\label{eq:n_sep}
\end{equation}

\noindent One consequently obtains the following expression for the
effective chemical potential:

\begin{equation}
\mu _{n}=-s_{n}(A_{0}-A,Z_{0}-Z)+f^{*}.  \label{eq:n_binary}
\end{equation}
where the reduced free excitation energy has been approximated by its low
energy limit, 
\begin{equation}
f^{*}=-\frac{T^{2}}{\varepsilon _{0}},\text{ ~~}\varepsilon _{0}=8MeV.
\label{eq:free_e}
\end{equation}
For the decay $^{112}$Sn$\longrightarrow ^{12}$C$+$X , the chemical
potential at $T=0$, i.e., $-s_{n}(A_{0}-A,Z_{0}-Z)$, is plotted as the stars
in Fig.(\ref{fig:n_chem}); the binding energies for these calculations were
calculated using the liquid-drop parametrization in ref.\cite{sneppen}. The
reduced free energy $f^{*}$ gives a reasonable estimate for the trend with
excitation energy. The dot-dashed line in Fig.(\ref{fig:n_chem}) gives the
chemical potential predicted from Eq. (\ref{eq:n_binary}) for $%
E_{0}^{*}/A=3\;MeV$ ($T=4.58MeV$)$.$ The predicted trend is close to that
predicted by the SMM model (solid circles and squares) but has a somewhat
stronger dependence on $N-Z.$

In general, the slope of the effective neutron chemical potential is getting
slightly flatter as the excitation energy or temperature increases. If we
consider that the system undergoes a multiple fragment decay at higher
temperatures, it is clear that approximating the entropy of the remaining
system by that of a residue of comparable mass becomes rather inaccurate.
The constraints imposed on the total system by the isospin asymmetry of one
observed fragment should, in that case, be less significant. While there is
a mass dependence to the effective chemical potential that is inconsistent
with the grand canonical approach, it is useful to note that the mass
dependence of the chemical potential (for these systems of more than 100
nucleons) is small if one is mainly concerned with nuclei near the valley of
stability. If one cancels the chemical potential effects by constructing
double ratios like that of the Algergo formula, the consequence of such
finite size effects becomes negligible indeed.

\section{Influence of Secondary Decay\newline
}

\label{sec:sec_dec} As discussed in sect.\ \ref{sec:assumptions}, fragments
are formed in excited states as well as in their ground states,
corresponding to the breakup temperature. Fragments in short lived excited
states decay before they are detected and, therefore, the observed yields
differ from that of the primary fragments. The effects of secondary decay on
the isotopic yields and isotopic temperatures have already been reported by
some authors (see for example \cite{Xi96,Bondorf98,Xi98b}). Although the
approaches employed in the description of the decay of hot primary fragments
are different, all those works qualitatively agree on the point that the
isotopic temperature is lower than the thermodynamical one.

At the quantitative level, details of the population and decay of the
excited fragments are important. One issue concerns the importance of
utilizing empirical binding energies, energy levels and decay branching
ratios for the excited fragments. Fig. (\ref{Y_2nd}) shows the primary and
secondary carbon isotopic distributions for the decay of a $^{112}Sn$
nucleus at initial excitation energies of $E_{0}^{*}/A=4$ and $6\;MeV$. The
primary distribution (solid line) is calculated by considering empirical
binding energies within the SMM for hot fragments. The simplified Weisskopf
evaporative decay procedure of ref. \cite{Botvina87} is used for one final
distribution (dotted line). The other final distribution (dashed line) is
obtained by calculating the secondary decay for $Z\leq 10$ hot fragments, as
in ref. \cite{Xi98b,Nay92}, according to empirical nuclear structure
information regarding the excitation energies, spins, isospins and decay
branching ratios where available. For hot fragments with $Z\leq 10$ where
such information is not available, the decay is calculated according to the
Hauser-Feshbach formalism\cite{Haus52}. The contributions to this latter
calculation from the secondary decay of hot fragments with $Z>10$, are
calculated, for simplicity, via the secondary evaporative decay procedure of
ref. \cite{Botvina87}. Decays of fragments with $Z>10$ make a 15\%
contribution to the yields of $^{12}C$ isotopes that may be altered when the
decay of hot fragments with $Z>10$ is calculated more accurately.

Obviously, in Fig. (\ref{Y_2nd}), the final distribution after the empirical
secondary decay is much wider than the final distribution obtained via the
evaporative decay approach of ref. \cite{Botvina87}. This points out the
importance of using the empirical information in such calculations. This
also leads to the extraction of larger isotopic temperatures via Eq. (\ref
{eq:t_alb}) for the empirical approach. Temperatures for the Carbon isotope
thermometer and He-Li thermometer calculated for the two secondary decay
approaches are shown, for example, in Fig. (\ref{T_2nd}) for the
multifragmentation of a $^{112}Sn$ nucleus at $E_{0}^{*}/A=4-10$ MeV. For
reference, the curves $T_{MIC}$ and $T_{IMF}$ from Fig. (\ref{Thermo}) are
also shown as the dashed and solid lines in the figure. Clearly,
incorporating empirical information in the decay makes a significant
difference. Both calculations provide lower isotopic temperatures than have
been obtained in recent experiments \cite{Pochodzalla,huang,serfling,Xi98}.

It should be noted, however, that the simplified Weisskopf evaporative
decay, shown in Figs. (\ref{Y_2nd}) and (\ref{T_2nd}), is only used in the
SMM code of ref. \cite{Botvina87} to calculate the decay of fragments with $%
A>16.$ The decay of lighter fragments is calculated via a ''Fermi Breakup''
multiparticle decay formalism. This latter decay mechanism makes the
dominant contribution to the isotope temperatures calculated via the latter
SMM code in ref. \cite{lynen98}. Investigations of the experimental and
theoretical basis for the ''Fermi Breakup'' approach are needed, but are out
of the scope of the present work.

Regardless of the decay formalism, memory of the breakup stage is lost via
the secondary decay mechanism. The degree of memory loss depends on the
details of the secondary decay correction and on the role of short-lived
higher lying particle unbound states. A smaller degree of memory loss ensues
in models such as those of refs. \cite{gross,dasgupta,ees}, where few, if
any, particle unbound states are considered. The approach of ref. \cite
{Botvina87} represents the other extreme, wherein all states are considered
regardless of lifetime. This issue clearly needs further study to see
whether the role of particle unstable nuclei can be constrained, for
example, by direct measurements using techniques discussed in refs. \cite
{Nay92,marie} or by other experimental observables.

\section{Concluding Remarks\newline
}

\label{sec:conrem} We discussed some of main aspects that could cause
microcanonical predictions for isotopic distributions and isotopic
temperatures to differ from grand canonical calculations and influence the
determination of the breakup temperature and other experimental observables.
We investigate this problem by checking the consistency of the grand
canonical expression for the isotopic yields against the approximately
microcanonical SMM predictions and explore the potential role which may be
played by variations in the temperature and in the effective chemical
potentials. These variations occur as a consequence of the finite size of
the disintegrating system and are therefore present in all microcanonical
calculations.

Concerning the temperature variation, we find that this causes the isotopic
yields obtained with the approximately microcanonical SMM simulations for
the primary distribution to differ from those of the grand-canonical
ensemble by factors of order unity. One difference stems from the averaging
over the temperatures corresponding to the different breakup partitions.
These vary because the total binding, coulomb and translational kinetic
energies vary from partition to partition and by subtraction, the thermal
energy must vary as well. A simple and relatively accurate prescription that
accounts for these temperature variations was given that may also prove
useful for estimating the influence of thermal averaging over the variations
in the actual excitation energy deposition within a data set that is
constrained by an experimental cut on the estimated energy deposition.

We also extract effective chemical potentials by comparing approximate
microcanonical and grand canonical expressions for the isotopic yields.
These effective chemical potentials are approximately the same for isotopes
of different elements that lie along the valley of beta stability, but vary
as a function of (N-Z). For example, we observe for the neutron chemical
potential a dependence upon (N-Z) that can be understood at low excitation
energies to arise from the dependence of the neutron separation energy on
the location of the accompanying residue relative to the line of beta
stability.

Typically, these variations in temperature and effective chemical potential
cause variations in the isotopic yields of order unity. The logarithmic
relation between the isotopic temperature and the yields means that the
latter may be wrongly predicted by a factor of two and one may still find a
reasonable agreement between the approximate microcanonical and the isotopic
temperatures provided the binding energy difference $\Delta B$ is
significantly larger than the temperature. When the effects of secondary
decay is taken into account, however, the yields can change by more than an
order of magnitude and the temperature values can decrease appreciably.
While the magnitude of this change is not yet unambiguously established, it
was shown that the incorporation of empirical information about the decay is
essential for quantitative comparisons to experimental data. Measurements
that quantify the role of higher lying particle unstable states are
essential for determining the magnitude of these secondary decay corrections.

\acknowledgments

We would like to acknowledge important discussions with Alexander Botvina
and HongFei Xi during the early stages of this work. R.D. and S.R.S. thank
the NSCL at MSU for support during visits on which part of this work was
performed, and also the MCT/FINEP/CNPq (PRONEX) program, under contract
\#41.96.0886.00, CNPq, FAPERJ, and FUJB for partial financial support. This
work was supported in part by the National Science Foundation under Grant
No. PHY-95-28844.

\begin{figure}[tbp]
\caption{The points denote distributions of temperatures calculated with the
SMM approach for the decay of a $^{112}Sn$ nucleus at three different
excitation energies. The lines denote gaussian fits to the calculated
distributions.}
\label{Tdist}
\end{figure}

\begin{figure}[tbp]
\caption{The points denote temperature distributions calculated with the SMM
approach for the different isotopes considered in the carbon thermometer for
an excitation energy of $E_{0}^{*}/A=6$MeV. The lines denote gaussian fits
to the calculated distributions.}
\label{Tdist_i}
\end{figure}

\begin{figure}[tbp]
\caption{Comparisons of various primary temperatures $T_{MIC}$, $T_{IMF}$
and $T_{iso}^{smm}$ from the SMM and $T_{iso}^{cal}$ from the analytical
calculation in the grand canonical limit. For details see the text. One
point is missing for $T_{iso}^{cal}$ with $\sigma_{T}=0.8$MeV because the
calculated value for p for the correction term in Eq.(\ref{eq:aveY}) becomes
negative at $E_{0}^{*}/A=3$MeV, i.e. the expansion breaks down in this case.}
\label{Thermo}
\end{figure}

\begin{figure}[tbp]
\caption{The solid squares and circles denote the free proton and neutron
yields, respectively, calculated via the SMM approach. The solid and dashed
lines denote fits to the calculated yields following Eq.\ (\ref{temp_expa}).}
\label{PN_fit}
\end{figure}

\begin{figure}[tbp]
\caption{The squares, circles and triangles denote neutron chemical
potentials derived from Eq.\ (\ref{eq:n_effchem}) using SMM predictions for
Carbon and Lithium isotopic yields at various initial excitation energies
for the decay of the nucleus $^{112}Sn$. The stars and the dot-dashed line
denote approximate values calculated from Eq.(\ref{eq:n_binary}) for T=0 and
4.58 MeV, respectively. The error bars denote the statistical errors in the
calculation, which in many cases are too small to be observed in the figure.}
\label{fig:n_chem}
\end{figure}

\begin{figure}[tbp]
\caption{Primary (solid line) and final Carbon isotopic distributions
calculated for the decay of the nucleus $^{112}Sn$ using (dashed line) and
neglecting (dotted line) the empirical nuclear structure information in the
secondary decay process. The error bars denote the statistical errors in the
calculation, which in many cases are too small to be observed in the figure.}
\label{Y_2nd}
\end{figure}

\begin{figure}[tbp]
\caption{Isotopic temperatures for Carbon and He-Li thermometers calculated
with the SMM model for the decay of the nucleus $^{112}Sn$ using (solid
symbols) and neglecting (open symbols) the empirical nuclear structure
information in the secondary decay process. The lines are the same as those
shown in Fig. (\ref{Thermo}). The error bars denote the statistical errors
in the calculation, which in many cases are too small to be observed in the
figure.}
\label{T_2nd}
\end{figure}


\begin{references}
\bibitem{foot1}  Some of the approximations invoked in ref. \cite{Botvina87}
are also found in ref. \cite{Bonrep}. Readers should therefore examine the
papers in ref. \cite{Bondorf} rather than ref. \cite{Bonrep} for further
details concerning the SMM approach adopted in this paper.
\bibitem{foot}  To be consistent with the thermodynamical expressions for
the free energy we handle the term involving the spin degeneracy factor $%
g_{AZ}^{gs}$ differently than it is handled in ref. \cite{Bondorf}. Within
the approximation in ref. \cite{Bondorf} of $g_{AZ}^{gs}=1,$ this term
vanishes.
\end{references}
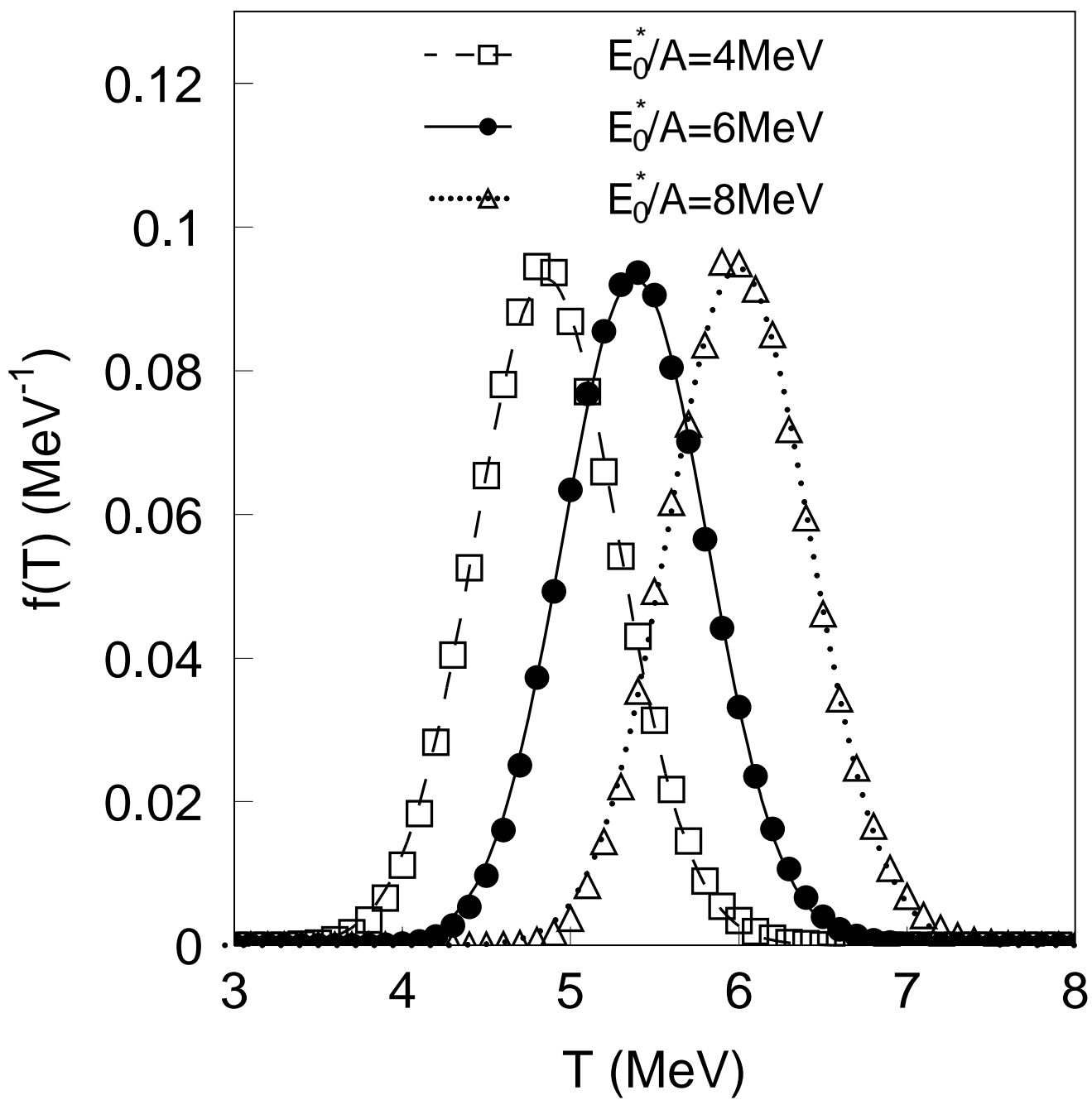
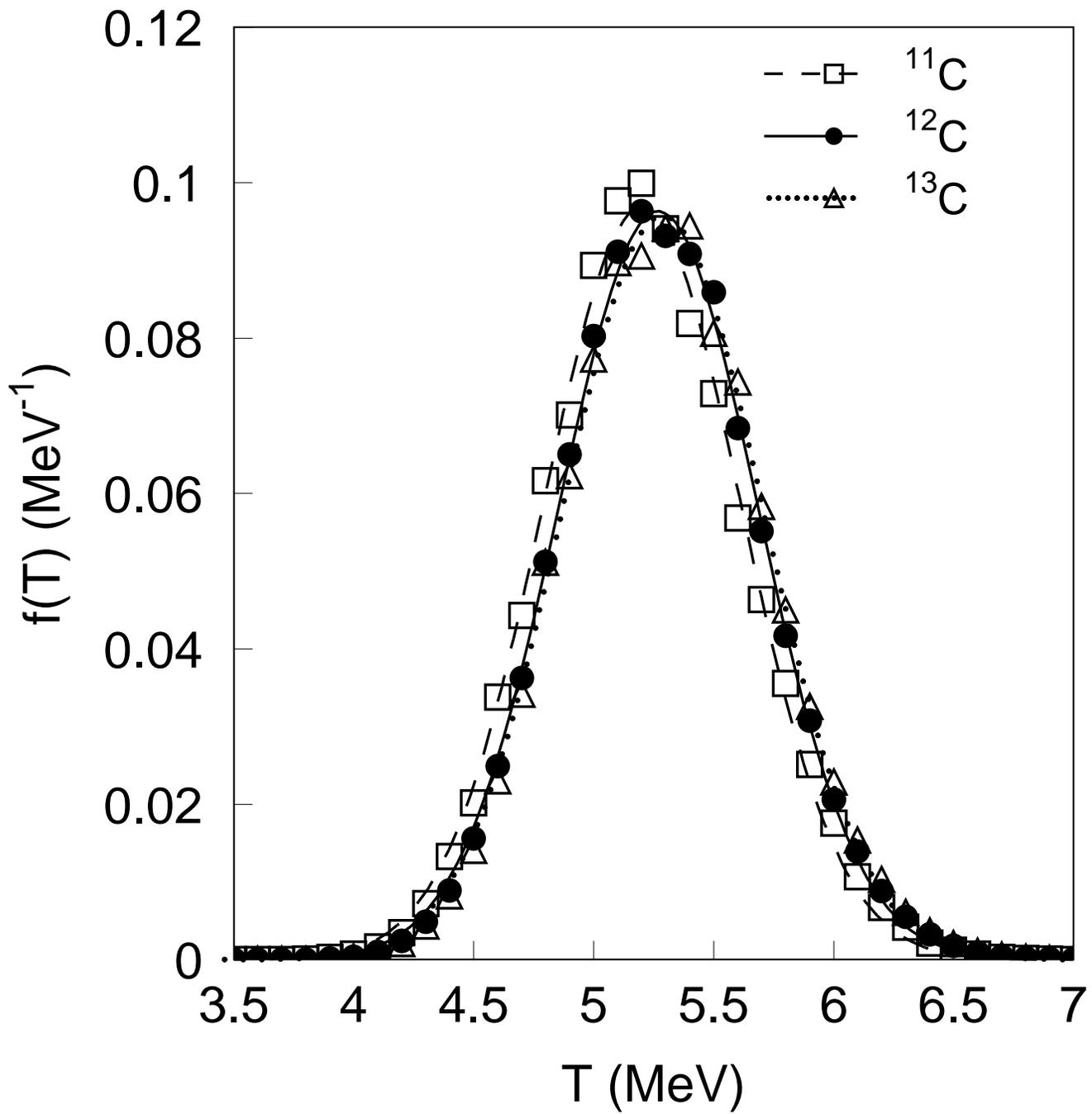
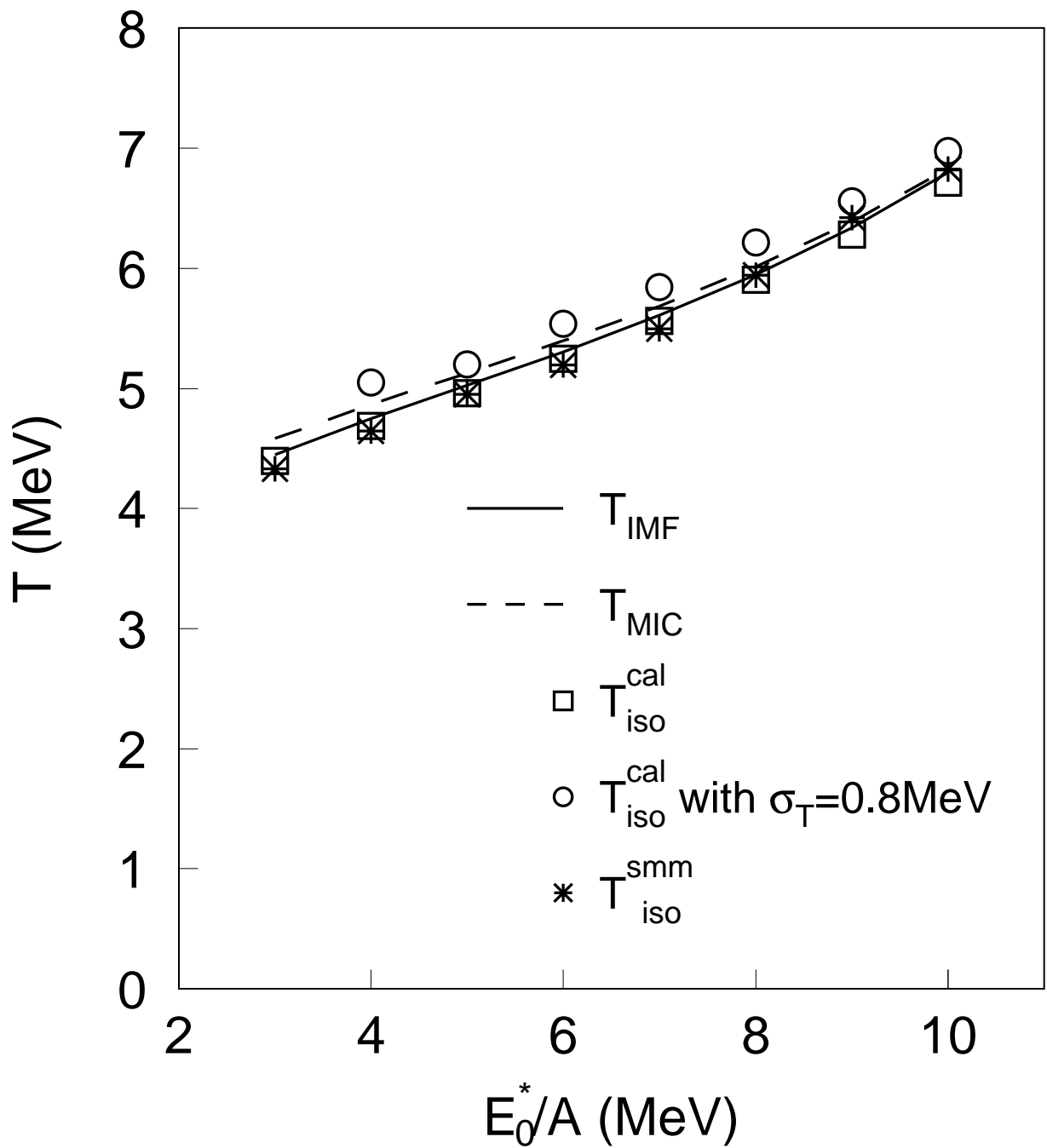
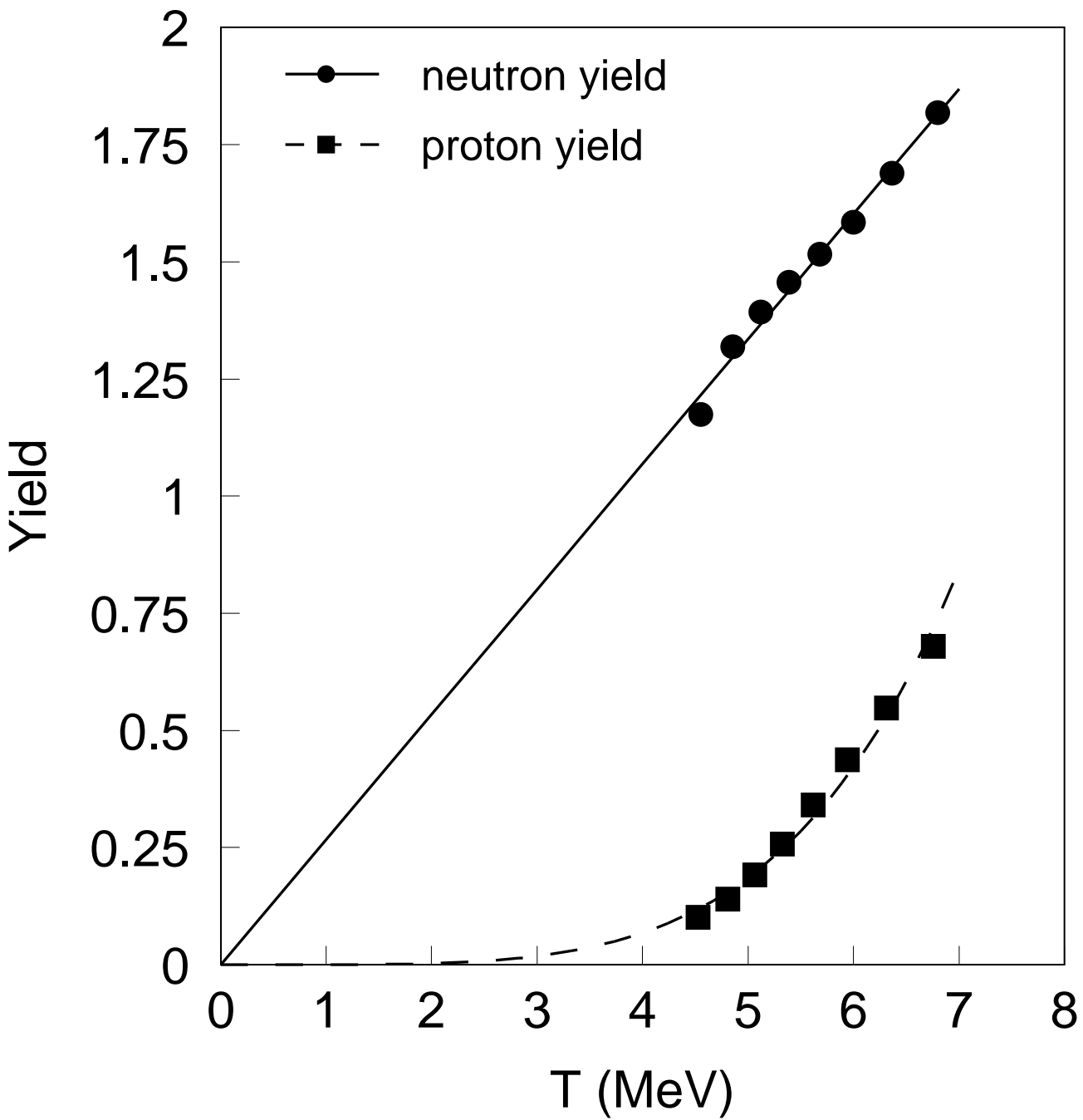
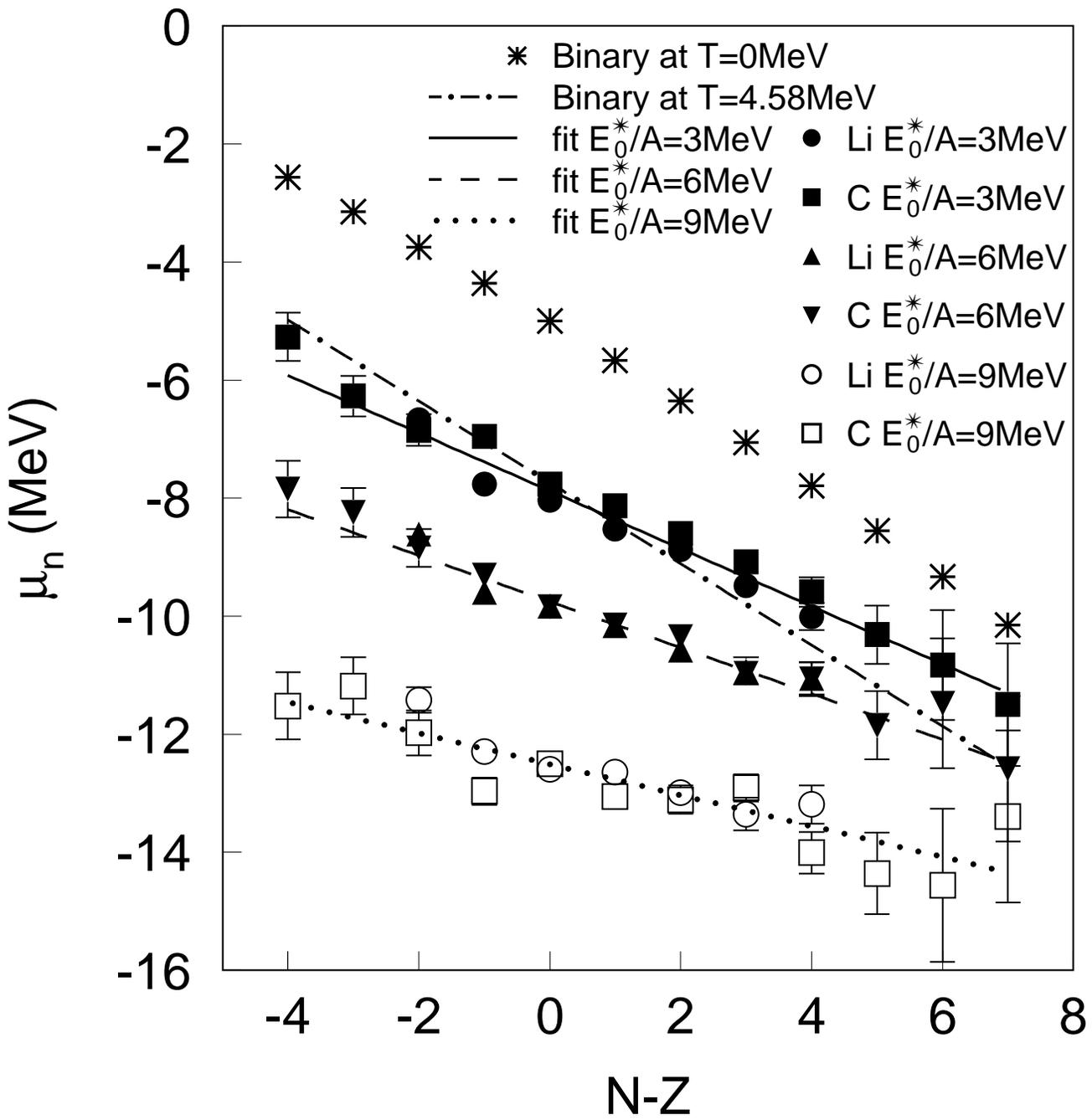
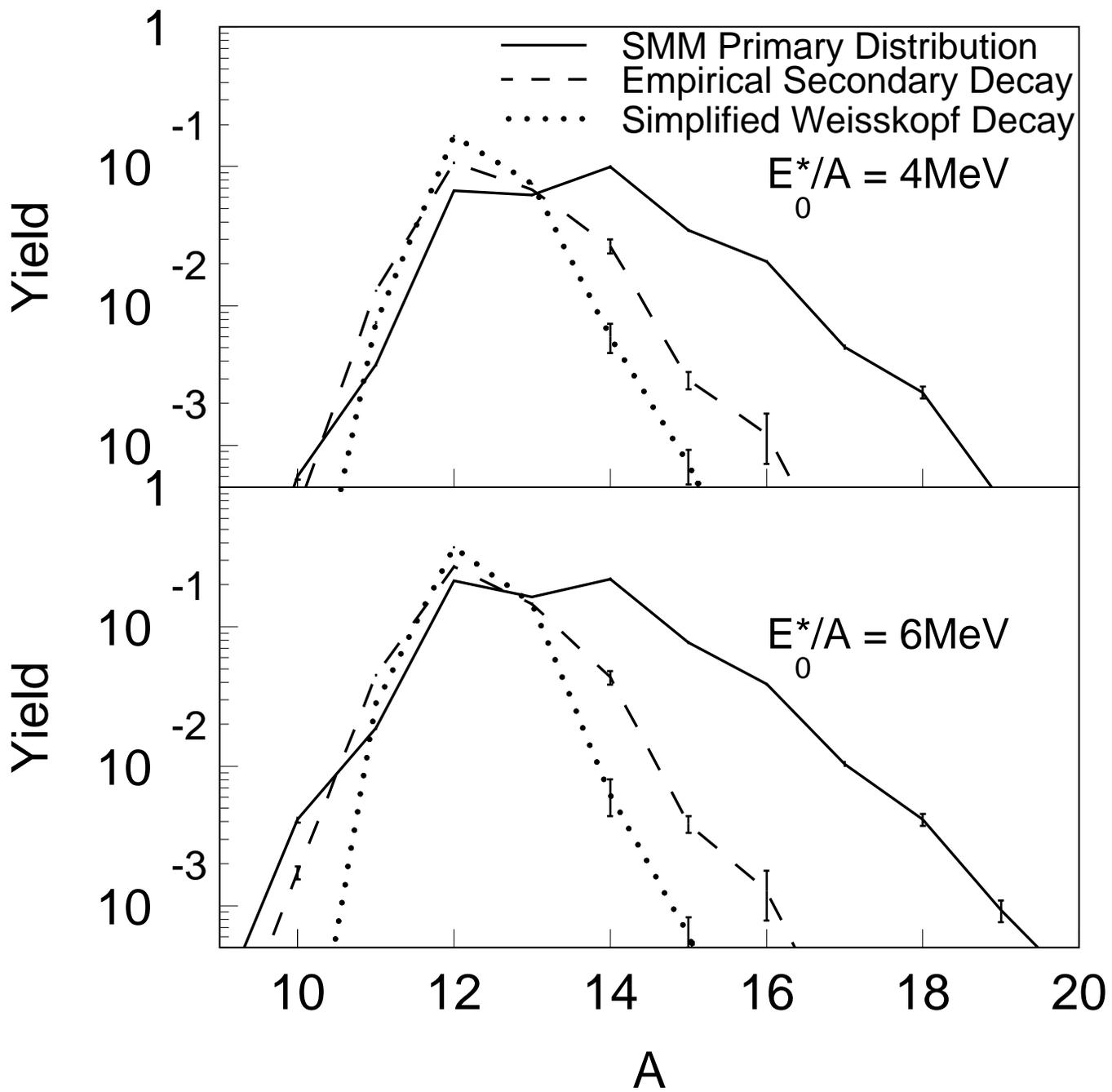
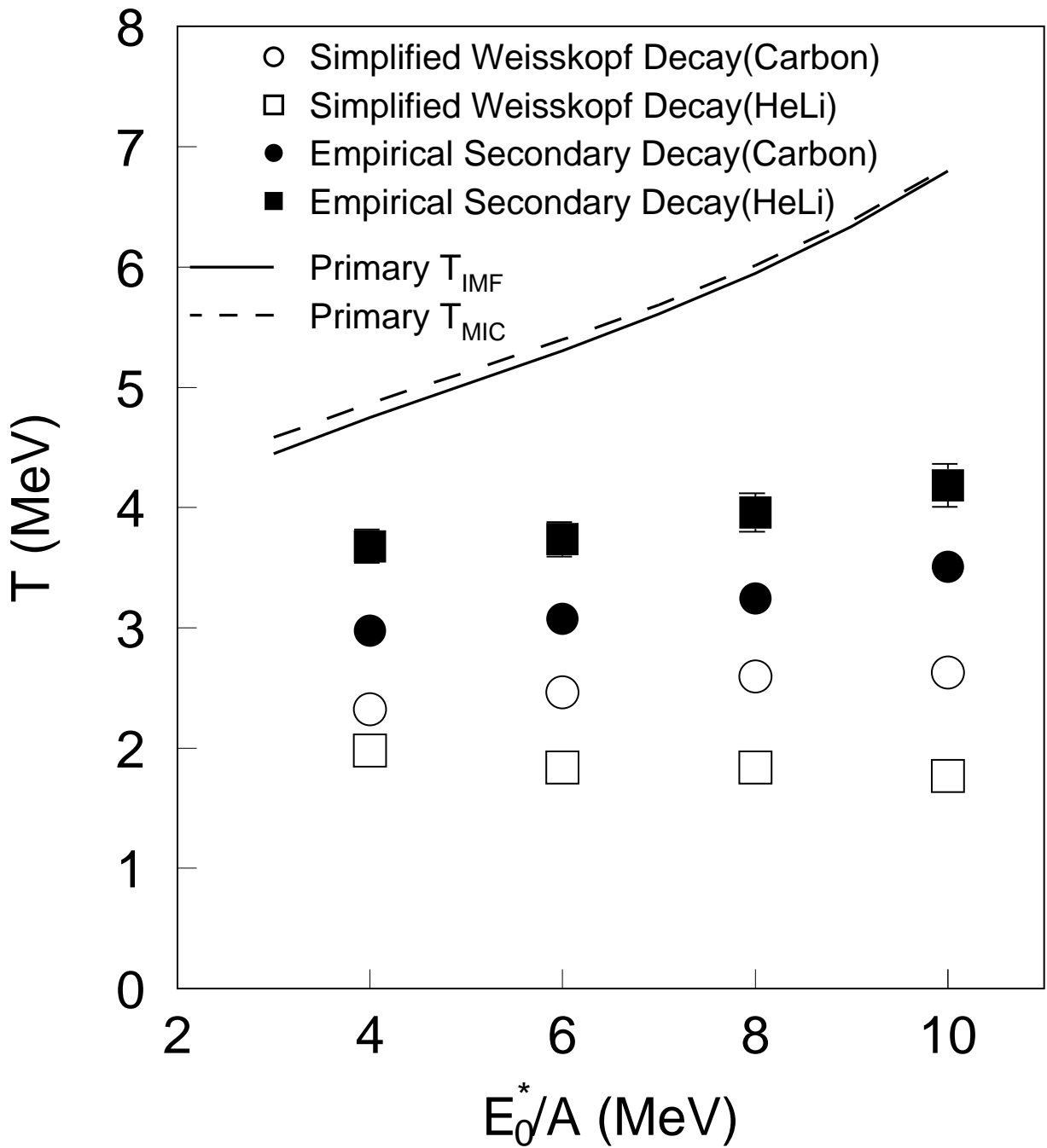
\end{document}